\documentclass[preprint,showpacs,preprintnumbers,amsmath,amssymb]{revtex4}
\usepackage{graphicx}
\usepackage{dcolumn}
\usepackage{bm}

\begin{document}


\title{Exposure of Nuclear Track Emulsion to a Mixed Beam of Relativistic $^{12}$N, $^{10}$C, and $^7$Be Nuclei}

\author{R.~R.~Kattabekov}
   \affiliation{Joint Insitute for Nuclear Research, Dubna, Russia}  
\author{K.~Z.~Mamatkulov}
   \affiliation{Joint Insitute for Nuclear Research, Dubna, Russia}  
\author{D.~A.~Artemenkov}
   \affiliation{Joint Insitute for Nuclear Research, Dubna, Russia}
\author{V.~Bradnova}
   \affiliation{Joint Insitute for Nuclear Research, Dubna, Russia}
\author{D.~M.~Zhomurodov}
   \affiliation{Joint Insitute for Nuclear Research, Dubna, Russia}
\author{M.~Haiduc}
   \affiliation{Institute of Space Sciences, Magurele, Romania}
\author{S.~P.~Kharlamov}
   \affiliation{Lebedev Institute of Physics, Russian Academy of Sciences, Moscow, Russia}
\author{V.~N.~Kondratieva}
   \affiliation{Joint Insitute for Nuclear Research, Dubna, Russia}
\author{D.~O.~Krivenkov}
   \affiliation{Joint Insitute for Nuclear Research, Dubna, Russia}  
\author{A.~I.~Malakhov}
   \affiliation{Joint Insitute for Nuclear Research, Dubna, Russia}
\author{G.~I.~Orlova}
   \affiliation{Lebedev Institute of Physics, Russian Academy of Sciences, Moscow, Russia}
\author{N.~G.~Peresadko}
   \affiliation{Lebedev Institute of Physics, Russian Academy of Sciences, Moscow, Russia}
\author{Z.~A.~Igamkulov}
   \affiliation{Joint Insitute for Nuclear Research, Dubna, Russia}
\author{N.~G.~Polukhina}
   \affiliation{Lebedev Institute of Physics, Russian Academy of Sciences, Moscow, Russia}
\author{P.~A.~Rukoyatkin}
   \affiliation{Joint Insitute for Nuclear Research, Dubna, Russia}
\author{V.~V.~Rusakova}
   \affiliation{Joint Insitute for Nuclear Research, Dubna, Russia}
\author{R.~Stanoeva}
   \affiliation{Joint Insitute for Nuclear Research, Dubna, Russia}
\author{S.~Vok\'al}
   \affiliation{P. J. \u Saf\u arik University, Ko\u sice, Slovak Republic}
\author{P.~I.~Zarubin}
     \email{zarubin@lhe.jinr.ru}
     \homepage{http://becquerel.jinr.ru}
   \affiliation{Joint Insitute for Nuclear Research, Dubna, Russia}
\author{I.~G.~Zarubina}
   \affiliation{Joint Insitute for Nuclear Research, Dubna, Russia}
   
\date{\today}

\begin{abstract}
\indent 
A nuclear track emulsion was exposed to a mixed beam of relativistic $^{12}$N, $^{10}$C, and $^7$Be nuclei having a momentum of 2 GeV/$c$ per nucleon. The beam was formed upon charge exchange processes involving $^{12}$C primary nuclei and their fragmentation. An analysis indicates that $^{10}$C nuclei are dominant in the beam and that $^{12}$N nuclei are present in it. The charge topology of relativistic fragments in the coherent dissociation of these nuclei is presented.\par
\indent \par
\indent DOI: 10.1134/S1063778810120161\par
\end{abstract}

\maketitle

\section{\label{sec:level1}INTRODUCTION}

\indent The use of accelerated nuclei, including radioactive ones, makes it possible to extend qualitatively the spectroscopy of cluster systems and to render it more versatile. A configuration overlap of the ground state of an accelerated nucleus with final cluster states manifests itself most strongly in the case of dissociation occurring at the periphery of the target nucleus and involving excitation transfer in the vicinity of cluster-binding thresholds. Owing to the collimation of projectile fragments, a determination of interactions as peripheral ones becomes easier as one moves toward energies in excess of 1 GeV per nucleon. The thresholds for their detection disappear, and the energy lost by fragments in a detector material is minimal. Thus, qualitatively new possibilities for studying cluster systems arise in the relativistic region in relation to the region of low energies. The method of nuclear track emulsions ensures the possibility of observing in minute detail and with a record spatial resolution multiparticle systems of relativistic fragments formed as dissociation products.\par
\indent Events of a coherent dissociation of nuclei, which proceeds without the formation of target fragments and mesons, into narrow jets of light and extremely light nuclei whose total charge is close to the charge of the primary nucleus constitute a moderately small fraction of observed interactions. The most peripheral of them are not accompanied by the formation of target fragments \cite{Andreeva} (so-called \lq\lq white\rq\rq~stars). The distributions of probabilities for cluster configurations manifest themselves in coherent-dissociation processes, and these distributions are peculiar to each individual nucleus involved.\par
\indent A series of runs of irradiation of a nuclear track emulsion according to the BECQUEREL Project \cite{web1} at the nuclotron of the Joint Institute for Nuclear Research (JINR, Dubna) with beams of a family of cluster nuclei $^{7,9}$Be, $^{8,10,11}$B, $^{9,10}$C, and $^{12,14}$N \cite{Andreeva,Peresadko,Artemenkov,Stanoeva,Karabova,Krivenkov,Rukoyatkin,Shchedrina} continued runs of irradiation at the JINR synchrophasotron, which were initially performed by using the cluster nuclei $^{12}$C \cite{Belaga} and $^6$Li \cite{Adamovich}. This created preconditions for analyzing various ensembles of light and extremely light nuclei detected under identical observation conditions. Knowledge of the distributions of the charge topology of \lq\lq white\rq\rq~stars in a nuclear track emulsion turned out to be of practical importance in estimating the composition of secondary beams of nuclei and made it possible, for example, to prove the dominance of $^8$B \cite{Stanoeva} and $^9$C \cite{Krivenkov} nuclei. Investigations of coherent dissociation of neutron-deficient nuclei are advantageous since they ensure the most comprehensive observation. In the present study, the approach developed and tested in the aforementioned investigations is used to explore the cluster features of the $^{12}$N and $^{10}$C nuclei. The $^{10}$C nucleus is the only example of a system that possesses super-Borromean properties, since the removal from it of one of the four clusters in the 2$\alpha$~+~2$p$ structure (threshold of 3.8 MeV) leads to an unbound state. A feature peculiar to the $^{12}$N nucleus is that the proton separation energy is small in it (600 keV). For \lq\lq white\rq\rq~stars produced by $^{12}$N, it would therefore be natural to expect the leading role of the $^{11}$C~+~$p$ channel. The $\alpha$~+~$^8$B (threshold of 8 MeV) and $p$~+~$^7$Be~+~$\alpha$ coherent-dissociation channels, as well as more complicated configurations leading to the cluster dissociation of cores in the form of the $^8$B and $^7$Be nuclei, are also possible.\par

\section{\label{sec:level2}DESCRIPTION OF THE EXPERIMENT}

\indent The production of $^{12}$N and $^{10}$C nuclei is possible in charge-exchange reactions involving accelerated nuclei of $^{12}$C and in their fragmentation. For the $^{10}$C and $^{12}$N nuclei, the ratios of the charges to the weights, Z$_{pr}$/A$_{pr}$, differ only by 3\%, and the momentum acceptance of the separating nuclotron channel is 2 to 3\% \cite{Rukoyatkin}. In view of this, the separation of these nuclei is impossible, so that $^{10}$C and $^{12}$N nuclei are present in the beam, forming a so-called beam cocktail. The contribution of $^{12}$N nuclei is much smaller than the contribution of $^{10}$C nuclei according to the ratio of the charge-exchange and fragmentation cross sections. The beam also contains $^7$Be nuclei, which differ from $^{12}$N in Z$_{pr}$/A$_{pr}$ by a value as small as 2\%. Because of the momentum spread, $^3$He nuclei may penetrate into the channel. For the neighboring nuclei $^8$B, $^9$C and $^{11}$C, the difference from $^{12}$N in Z$_{pr}$/A$_{pr}$ is about 10\%. Owing to this, they are suppressed in the irradiation of the track emulsion. The identification of $^{12}$N and $^7$Be nuclei in the irradiated track emulsion can be performed on the basis of beam-nucleus charges as determined by counting $\delta$ electrons along beam tracks. In the case of $^{10}$C, it is necessary to verify on the basis of the charge topology of \lq\lq white\rq\rq~stars whether the contribution of neighboring isotopes is small. These were arguments behind the proposal to expose a stack of nuclear track emulsion to a mixed beam of $^{12}$N, $^{10}$C, and $^7$Be nuclei.\par 	

\begin{figure}
    \includegraphics[width=4in]{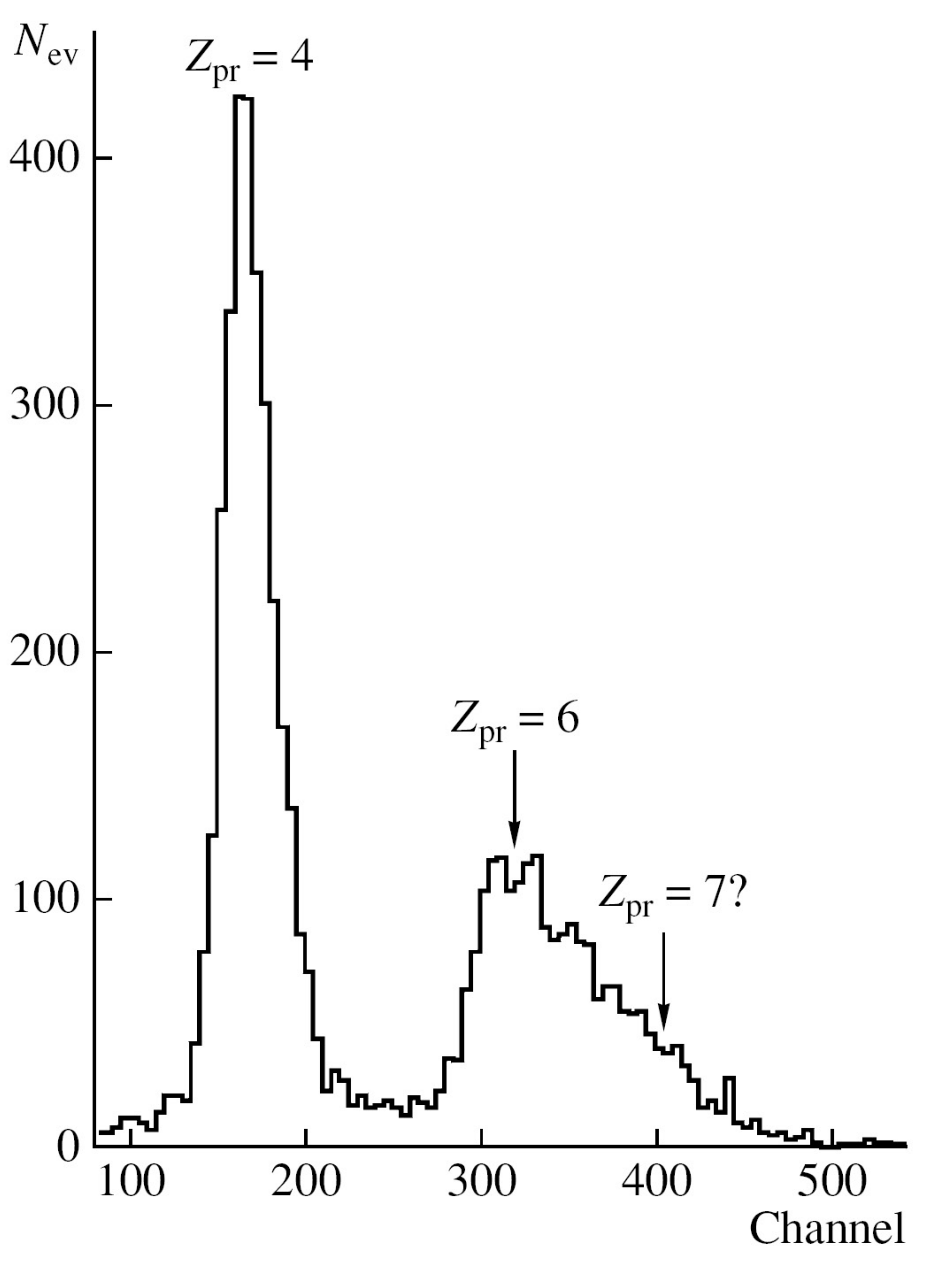}
    \caption{\label{Fig:1} Amplitude spectrum from a scintillation counter arranged at the locus of irradiation of the track-emulsion stack used, the beam-transportation channel being tuned to the separation of $^{12}$N nuclei; the peak positions are indicated for nuclei of charge Z$_{pr}$~=~4, 6, and 7.}
    \end{figure}
		
\indent A beam of $^{12}$C nuclei with a momentum of 2 GeV/$c$ per nucleon was accelerated at the JINR nuclotron and was extracted and directed to a generating target. Through a beam-transportation channel, including four deflecting magnets over a base 70 m long, a secondary beam characterized by a magnetic rigidity optimal for selecting $^{12}$N nuclei and by the same momentum per nucleon as $^{12}$C nuclei was delivered to the locus of irradiation of the trackemulsion stack used \cite{Rukoyatkin}. The amplitude spectrum from the scintillation counter arranged at this locus is indicative of the dominance of the isotopes $^3$He and $^7$Be and isotopes of C, the presence of an admixture of $^{12}$N nuclei, and an almost complete absence of $^8$B nuclei (Fig. 1). A stack of 15 layers of BR-2 nuclear track emulsion was exposed to a secondary beam of this composition.\par
\indent The initial step of scanning the track-emulsion layers consisted in visual searches for beam tracks of charge Z$_{pr}$~=~1,~2 and Z$_{pr}~>~$2. The ratio of the numbers of Z$_{pr}$~=~1,~2 and Z$_{pr}~>~$2 beam tracks was about 1~:~3~:~18. For the sake of comparison, we indicate that, in the case of irradiation with $^9$C nuclei, this ratio was about 1 : 10 : 1. Thus, the contribution of $^3$He nuclei in this run was much smaller, which improved radically the efficiency of irradiation and the rate of event searches.\par

\begin{figure}
    \includegraphics[width=5in]{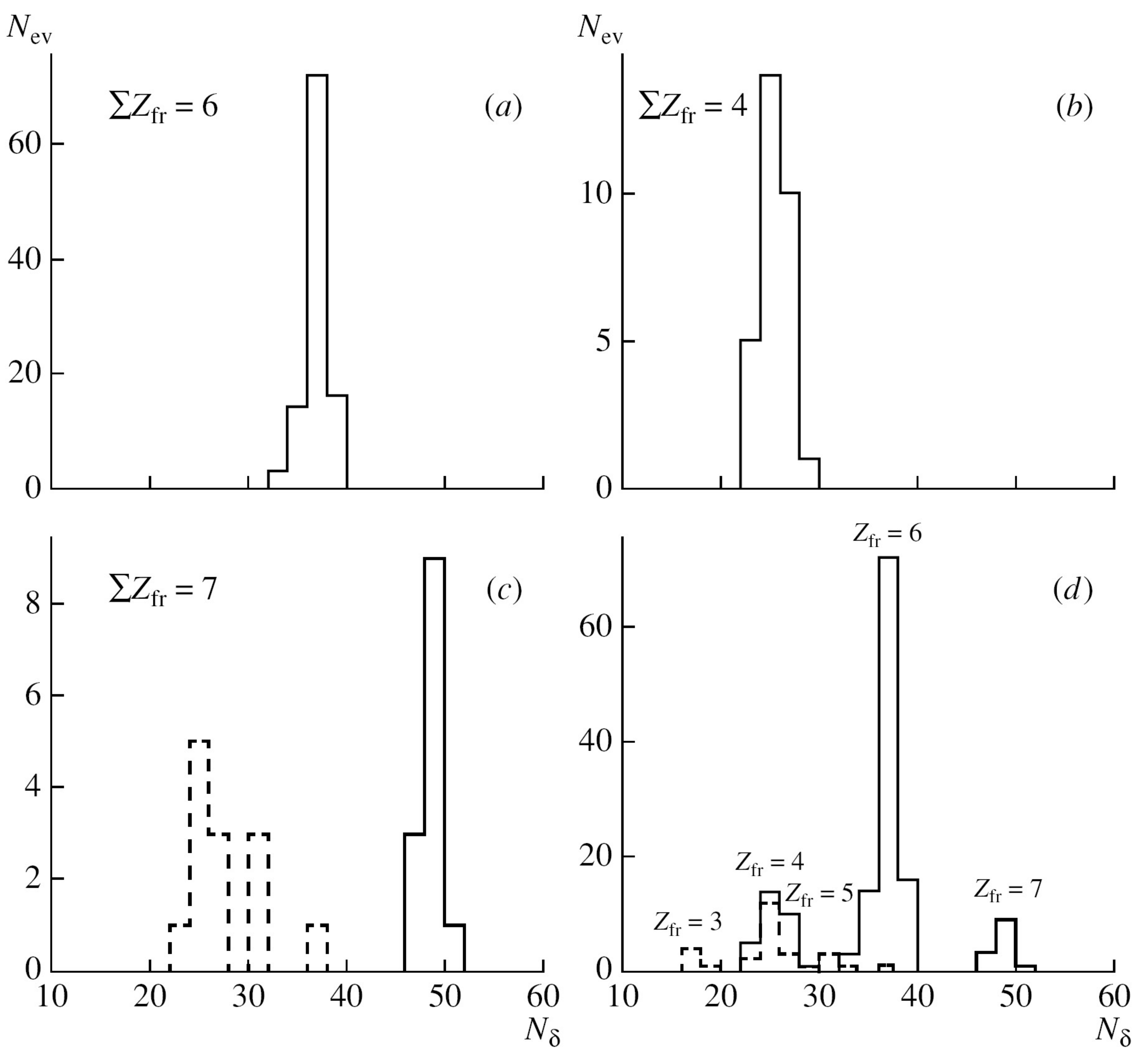}
    \caption{\label{Fig:2} Distribution of the number N$_{tr}$ of tracks of beam particles and secondary fragments (dashed-line histogram) with respect to the mean number of $\delta$ electrons, N$_{\delta}$, over 1 mm of the track length in ($a$) 2He~+~2H and ($b$) 2He and He~+~2H \lq\lq white\rq\rq~stars and in events featuring ($c$) Z$_{fr}~>~$2 fragments along with ($d$) the distribution of N$_{tr}$ with respect to N$_{\delta}$ for all measured events.}
    \end{figure}
	
\section{\label{sec:level3}CHARGE TOPOLOGY OF \lq\lq WHITE\rq\rq~STARS}

\indent Interactions in track-emulsion layers were sought on the basis of Z$_{pr}~>~$2 primary tracks without selections. The search for \lq\lq white\rq\rq~stars and charge measurements for them were performed in eight layers.\par
	
\begin{table}
\caption{\label{Table:1} Distribution of the number of \lq\lq white\rq\rq~stars N$_{ws}$ over dissociation channels for which the total charge of fragments is $\sum$Z$_{fr}$~=~6}
\label{Table:1}       
\begin{tabular}{l|c|c|c|c|c}
\hline\noalign{\smallskip}
Channel & 2He + 2H & He + 4H & $^8$B + H & $^7$Be + He & $^7$Be + 2H \\
\noalign{\smallskip}\hline\noalign{\smallskip}
N$_{ws}$ & 95 & 14 & 1 & 5 & 3 \\
\noalign{\smallskip}\hline
\end{tabular}
\end{table}

\indent The scanning along the total length of primary tracks that was equal to 462.6 m revealed 3258 inelastic
interactions, including 330 \lq\lq white\rq\rq~stars that involve only He and H relativistic fragments and 27 \lq\lq white\rq\rq~stars that involved Z$_{fr}~>~$2 fragments. The angular fragmentation cone was specified by the soft condition $\theta_{fr}~<~8{^\circ}$.\par

\begin{table}
\caption{\label{Table:2} Distribution of the number of \lq\lq white\rq\rq~stars N$_{ws}$ over dissociation channels for which the total charge of fragments is $\sum$Z$_{fr}$~=~7, with the measured charge of the beam track being Z$_{pr}$~=~7}
\label{Table:2}       
\begin{tabular}{l|c|c|c|c|c|c|c}
\hline\noalign{\smallskip}
Channel & $^7$Be + 3H & $^8$B + 2H & C + H & 2He + 3H & He + 5H & He + H & $^7$Be + He + H \\
\noalign{\smallskip}\hline\noalign{\smallskip}
N$_{ws}$ & 4 & 3 & 1 & 6 & 3 & 2 & 2 \\
\noalign{\smallskip}\hline
\end{tabular}
\end{table}

\indent Because of the presence of Z$_{fr}~>~$2 fragments, it was necessary to perform a charge identification of beam (Z$_{pr}$) and secondary (Z$_{fr}$) tracks. In order to calibrate this procedure, we measured the mean density of $\delta$ electrons, N$_{\delta}$, over 1 mm of length along the tracks of beam nuclei producing 2He~+~2H, 2He, and He~+~2H \lq\lq white\rq\rq~stars, as well as in stars featuring Z$_{fr}~>~$2 fragments as candidates for $^{12}$N (see Fig. 2). We observed a correlation of the charge topology $\sum$Z$_{fr}$ and N$_{\delta}$, and this enabled us to determine Z$_{pr}$ for each beam track. In this way, we performed a calibration that made it possible to determine the charges of Z$_{fr}~>~$2 fragments by the same method. The summed distribution for those measurements is presented in Fig. 2$d$. This spectrum is indicative of the presence of $^{12}$N nuclei in the composition of the beam.\par
\indent For \lq\lq white\rq\rq~stars for which the conditions Z$_{pr}$~=~$\sum$Z$_{fr}$ and $\sum$Z$_{fr}$~=~6 hold, Table 1 shows the distribution of their number N$_{ws}$ over dissociation channels. For the case of $\sum$Z$_{fr}$~=~6, this condition was tested only within two layers in performing calibration, since there was no need for a complete test in view of the dominance of carbon nuclei. As might have been expected for the isotope $^{10}$C, themost probable channel was represented by 91 2He~+~2H events. The He~+~4H channel proved to be suppressed. Indeed, a peripheral dissociation of $^{10}$C requires overcoming a high threshold for the breakup of the alpha-particle cluster.\par
\indent For 20 events found here that are characterized by Z$_{pr}$~=~7 and $\sum$Z$_{fr}$~=~7 and which are associated with the dissociation of $^{12}$N nuclei, the distribution with respect to the charge topology is presented in Table 2. About half of the events feature a Z$_{fr}~>~$2 fragment, and this is markedly different from what we have for the case of the $^{14}$N \cite{Shchedrina} and $^{10}$C nuclei.\par

\section{\label{sec:level3}PRODUCTION OF UNSTABLE NUCLEI $^8$Be AND $^9$B}

\indent The unstable nucleus $^8$Be plays the role of a core in the structure of the $^{10}$C nucleus, and this must manifest itself in the dissociation process $^{10}$C~$\rightarrow$~$^8$Be. The relativistic-nucleus decays $^8$Be~$\rightarrow$~2$\alpha$ through the 0$^+$ ground state are identified by the appearance of alpha-particle pairs in the characteristic region of smallest divergence angles $\Theta_{2\alpha}$, which, at the momentum of 2 GeV/$c$ per nucleon, is bounded by the condition $\Theta_{2\alpha}~<~$10.5 mrad \cite{Artemenkov}. The excitation energy of an alpha-particle pair, $Q_{2\alpha}$~=~$M^*_{2\alpha}$~-~$M_{2\alpha}$, where $M^*_{2\alpha}$ is the invariant mass of the system of fragments, $M^*_{2\alpha}$~=~($\sum P_j$)$^2$~=~$\sum$($P_i~\times~P_k$), $P_{i,k}$ stand for the fragment-$i$ and fragment-$k$ 4-momenta as determined in the approximation of conservation of the primary momentum per nucleon, and $M_{2\alpha}$ is the doubled alpha-particle mass, has a physical meaning.\par

\begin{figure}
    \includegraphics[width=4in]{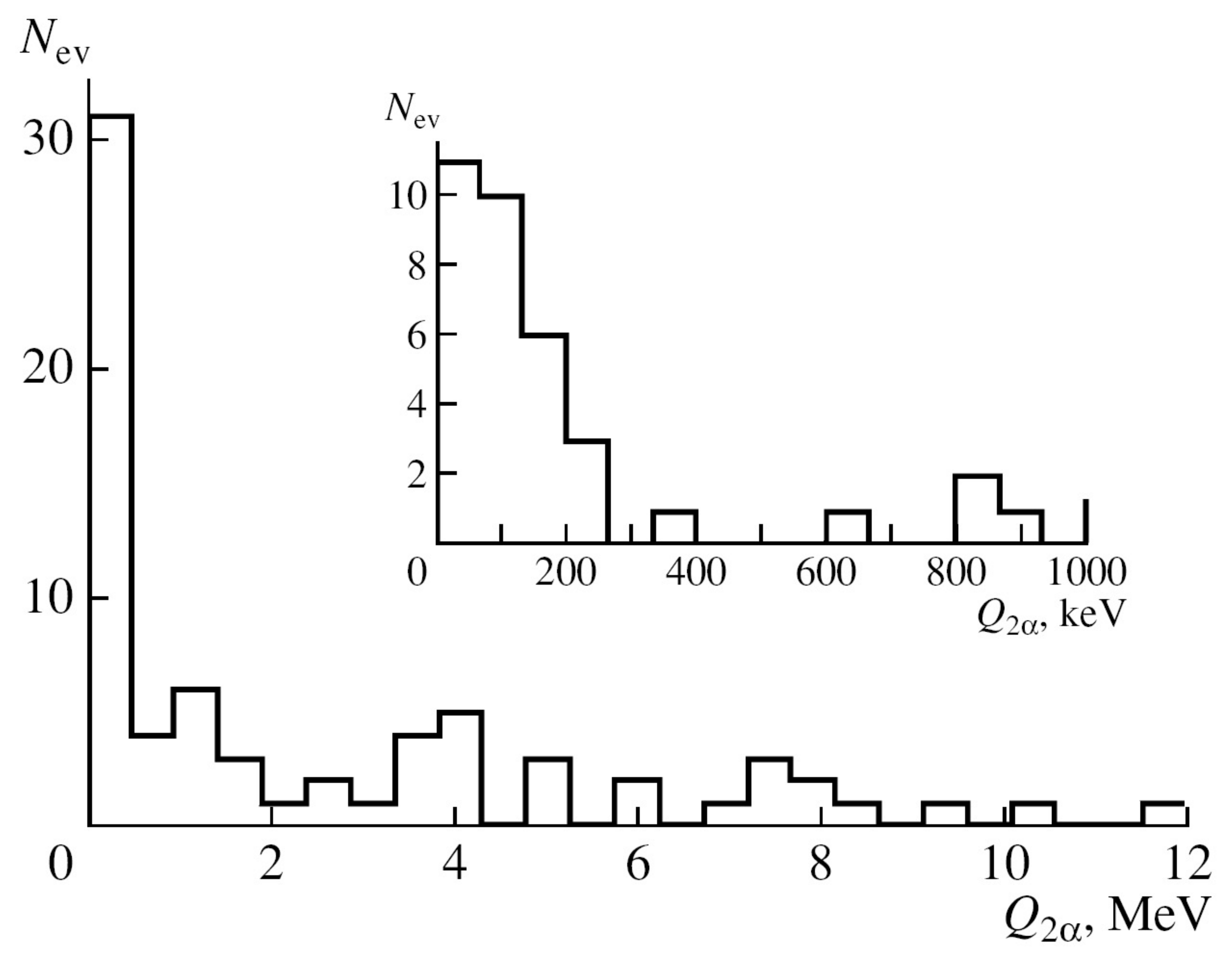}
    \caption{\label{Fig:3} Distribution of the number of \lq\lq white\rq\rq~stars, N$_{ws}$, having the 2He~+~2H topology with respect to the excitation energy $Q_{2\alpha p}$ of alpha-particle pairs. The inset shows the distribution $Q_{2\alpha p}$ on an enlarged scale.}
    \end{figure}

\indent The excitation-energy ($Q_{2\alpha}$) distribution of alphaparticle pairs from 91 2He~+~2H \lq\lq white\rq\rq~stars is shown in Fig. 3. Among them, 30 events are those in which $Q_{2\alpha}$ does not exceed 500 keV (see inset in Fig. 3). The average value $<Q_{2\alpha}>$ is 110~$\pm$~20 keV, and the root-mean-square scatter is $\sigma$~$\approx$~40 keV; this corresponds to the decays of the ground state of the $^8$Be nucleus \cite{Artemenkov}. The branching fraction of these decays corresponds to the cases of neighboring cluster nuclei.\par
	
\begin{figure}
    \includegraphics[width=4in]{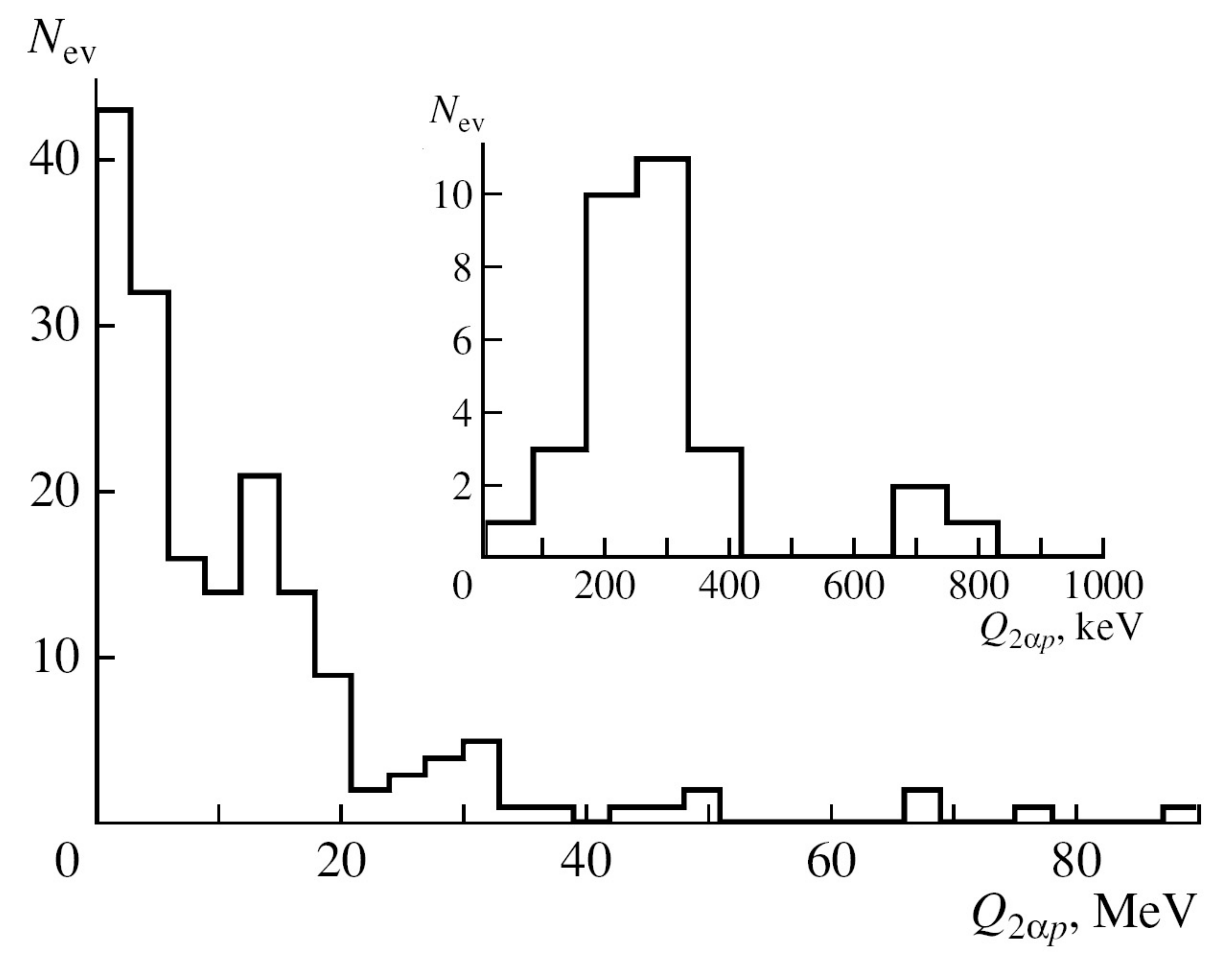}
    \caption{\label{Fig:4} Distribution of the number of \lq\lq white\rq\rq~stars, N$_{ws}$, having the 2He~+~2H topology with respect to the excitation energy $Q_{2\alpha p}$ of the 2$\alpha$~+~$p$ three-particle system. The inset shows the distribution $Q_{2\alpha p}$ on an enlarged scale.}
    \end{figure}

\indent The unstable nucleus $^9$B must be yet another product of the coherent dissociation of the $^{10}$C nucleus. Figure 4 shows the distribution of 2He~+~2H \lq\lq white\rq\rq~stars with respect to the excitation energy $Q_{2\alpha p}$ determined from the difference of the invariant mass of the 2$\alpha$~+~$p$ three-fragment system and the sum of the proton mass and the doubled alpha-particle mass. In 27 events, the value of $Q_{2\alpha p}$ for one of the two combinatorially possible three-particle systems $\alpha$~+~$\alpha$~+~$p$ does not exceed 500 keV either (see inset in Fig. 4). The average value $<Q_{2\alpha p}>$ is 250~$\pm$~15~keV, and the root-mean-square scatter is $\sigma$~=~74~keV. These values correspond to the decay of the ground state of the $^9$B nucleus through the $p$~+~$^8$Be~(0$^+$) channel with the energy and width values known to be, respectively, 185 keV and 0.54~$\pm$~0.21~keV \cite{web2}. In the distributions of $Q_{2\alpha}~<~$1~MeV and $Q_{2\alpha p}~<~$1~MeV, there is a clearcut correlation in the production of $^8$Be and $^9$B in the ground states. The appearance of one 2$\alpha$~+~2$p$ event for the $Q_{2\alpha p}$ values of 0.23 and 0.15 keV is noteworthy—that is, the two three-particle systems simultaneously correspond to the decay of the $^9$B nucleus. In all of the remaining cases of the production of a $^9$B nucleus, the second of the two possible values of $Q_{2\alpha p}$ is in excess of 500 keV.\par
\indent In addition, we have studied excitations of the $\alpha$~+~2$p$ system by using the statistical sample of 2He~+~2H \lq\lq white\rq\rq~stars that remains after the elimination of the decays of the $^9$B nucleus. In the spectrum of $Q_{\alpha2p}$, there is no explicit signal from the decays of the ground and first excited states of the unstable nucleus $^6$Be \cite{web2} — an estimate of its contribution does not exceed 20\%. This aspect deserves a further analysis with allowance for angular correlations of protons.\par

\section{\label{sec:level4}CONCLUSIONS}

\indent By and large, it seems that the charge topology of the dissociation of the nuclei explored in the present study is not contradictory and that the trackemulsion irradiation performed in our experiment is promising both for enlarging the statistical sample of $^{12}$N and $^{10}$C \lq\lq white\rq\rq~stars and for analyzing them in detail. Even at the present stage of analysis, one can also draw some physical conclusions about clustering features of the $^{12}$N and $^{10}$C nuclei.\par
\indent In practical aspects, our analysis of angular correlations supports the conclusion that $^{10}$C nuclei are dominant in the beam. The production of $^8$Be in the dissociation of $^{10}$C nuclei has a cascade character: $^{10}$C~$\rightarrow$~$^9$B~$\rightarrow$~$^8$Be. There is no sizable contribution from the decay $^8$Be~$\rightarrow$~2$\alpha$ through the first excited state (2$^+$), and this distinguishes qualitatively the $^{10}$C nucleus from $^9$Be. In the case of the $^9$Be nucleus \cite{Artemenkov}, the contributions of the 0$^+$ and 2$^+$ states of the $^8$Be nucleus to the $^9$Be~$\rightarrow$~$^8$Be transition proved to be close and corresponded to the weights adopted for these states in the calculation of the magnetic moment of the $^9$Be nucleus on the basis of the $n$ - $^8$Be two-body model \cite{Parfenova1,Parfenova2}.\par
\indent One can assume that the 2$^+$ state of the $^8$Be nucleus does not contribute to the ground state of the $^{10}$C nucleus and that only the 0$^+$ extended state forms its basis \cite{Wiringa}. Paired protons may play the role of a covalent pair in the $^{10}$C molecule-like system controlled by the $\alpha$~+~2$p$~+~$\alpha$ two-center potential. Such assumptions will be verified by analyzing correlations in 2$p$, 2$\alpha$, and $\alpha p$ pairs and, later on, in more complex configurations for the $p$~+~$^9$B, 2$p$~+~$^8$Be, and $\alpha$~+~$^6$Be unstable nuclei.\par
\indent In order to investigate nucleon clustering in the $^{12}$N nucleus, it is necessary to enlarge the statistical sample used and to identify H and He isotopes by the method of multiple scattering. The $^{11}$C nucleus does not manifest itself as the core of the $^{12}$N nucleus. The absence of $^8$B~+~He events (the respective threshold is 8 MeV) seems unexpected. It is conceivable that a $^7$Be or a $^8$B nucleus appears as the core of the $^{12}$N nucleus and that the remaining nucleons do not form an alpha-particle cluster.\par

\begin{acknowledgments}
\indent This work was supported by the Russian Foundation for Basic Research (project nos. 96-1596423, 02-02-164-12a, 03-02-16134, 03-02-17079, 04-02-17151, 04-02-16593, and 09-02-9126-ST-a) and by the Agency for Science at theMinistry for Education of the Slovak Republic and Slovak Academy of Sciences (grants VEGA nos. 1/2007/05 and 1/0080/08), as well as by grants from the Plenipotentiaries of Bulgaria, the Slovak Republic, the Czech Republic, and Romania at the Joint Institute for Nuclear Research (JINR, Dubna) in 2002–2009. We are grateful to U.S. Salikhbaev, member of the Uzbek Academy of Sciences, as well as to Professor R.N. Bekmirzaev (Dzhizak State Pedagogical Institute, Dzhizak) and Dr. Sci. (Phys.–Math.) K. Olimov (Institute for Physics and Technology, Tashkent).\par
\end{acknowledgments}

\end{document}